\documentstyle[sprocl,epsfig]{article}

\def\Journal#1#2#3#4{{#1} {\bf #2}, #3 (#4)}

\def\NPA{{\em Nucl. Phys.} A}
\def\NPB{{\em Nucl. Phys.} B}
\def\PLB{{\em Phys. Lett.}  B}
\def\PRL{\em Phys. Rev. Lett.}
\def\PRC{{\em Phys. Rev.} C}
\def\PRD{{\em Phys. Rev.} D}
\def\EPJC{{\em Euro. Phys. J.} C}

\def\be{\begin{equation}}
\def\ee{\end{equation}}
\def\bea{\begin{eqnarray}}
\def\eea{\end{eqnarray}}

\begin{document}

\title{$N^*$ RESONANCES IN $e^+e^-$ COLLISIONS AT BEPC}

\author{B. S. ZOU, H. B. LI}
\author{Representing BES Collaboration}

\address{Institute of High Energy Physics, CAS, P.O.Box 918(4),\\
Beijing 100039, P.R.China\\E-mail: zoubs@zou.ihep.ac.cn}

\maketitle\abstracts{
$J/\Psi$ particles are abundantly produced at the Beijing
Electron-Positron Collider (BEPC).
The $J/\Psi$ decays provide an excellent place for studying $N^*$
resonances.
For $J/\Psi\to\bar NN\pi$ and $\bar NN\pi\pi$, the $\pi N$ and $\pi\pi N$
systems are limited to be pure isospin $1/2$ due to isospin conservation.
This is a big advantage in studying $N^*$ resonances from $J/\Psi$ decays,
compared with $\pi N$ and $\gamma N$ experiments which suffer difficulty
on the isospin decomposition of $1/2$ and $3/2$.
All other $N^*$ decay channels which are presently under investigation at
CEBAF(JLab, USA), ELSA(Bonn,Germany) and GRAAL(Grenoble, France) with
real photon or space-like virtual photon can also
be studied at BEPC complementally with the time-like virtual photon.
The process $J/\Psi\to\bar NN^*$ or $N\bar N^*$ provides a new way to
probe the internal structure of the $N^*$ resonances.
The recent results and outlook of our new $N^*$ program at BEPC are
presented.}

\section{Introduction}

To understand the internal quark-gluon structure of nucleon and its
excited states $N^*$'s is one of the most important tasks in nowadays
particle and nuclear physics. The main source of information for the
baryon internal structure is their mass spectrum, various production and
decay rates. Our present knowledge of this aspect came almost entirely
from the old generation of $\pi N$ experiments of more than twenty years
ago. Considering its importance for the understanding of the
nonperturbative QCD, a new generation of experiments on $N^*$ physics with
electromagnetic probes (real photon and space-like virtual photon) has
recently been started at new facilities such as CEBAF at JLAB, ELSA at
Bonn, GRAAL at Grenoble.

\begin{figure}[htbp]
\epsfig{figure=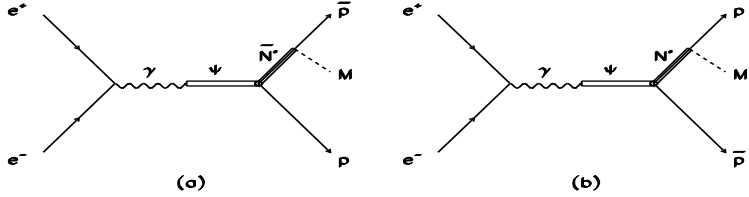,height=1.6in,width=4.7in}
\caption{Feynman graphs for $N^*$ and $\bar N^*$ production from
$e^+e^-$ collision through $J/\Psi$ meson.}
\label{figure:1}
\end{figure}

The $J/\Psi$ experiment at the Beijing Electron-Positron Collider (BEPC)
has long been known as the best place for looking for glueballs. But in
fact it is also an excellent place for studying $N^*$ resonances.
The corresponding Feynman graphs for the $N^*$ and $\bar N^*$ production
are shown in Fig.~\ref{figure:1}.

These graphs are almost identical to those describing the $N^*$
electro-production process if the direction of the time axis is
rotated by $90^o$. The only difference is that the virtual photon
here is time-like instead of space-like and couples to $NN^*$ through
a real vector charmonium meson $J/\Psi$. This fact leads to a few
advantages for studying $N^*$ resonances at the BEPC as the following.

\noindent (1) $J/\Psi\to N \bar N \pi$ and $N\bar N\pi\pi$ provide a
natural isospin $I=1/2$ filter for the $\pi N$ and $\pi\pi N$ systems due
to isospin conservation.

Although the existence of the $N^*(1440)$ and $N^*(1535)$ is
well-established, their properties, such as mass, width and decay
branching ratios etc., still suffer large experimental
uncertainties\cite{PDG}.
A big problem in extracting information on these $N^*$ resonances from
$\pi N$ and $\gamma N$ experiments is the isospin decomposition of 1/2 and
3/2 for $\pi N$ and $\pi\pi N$ systems\cite{Workman}. We expect that the
results from $J/\Psi$ decays will provide better determination of the
properties of these $N^*$ resonances.

\noindent (2) Interference between $N^*$ and $\bar N^*$ bands in
$J/\Psi\to N\bar N\pi$ Dalitz plots may help to distinguish some
ambiguities in the partial wave analysis of $\pi N$ two-body channel
alone from $\pi N$ and $\gamma N$ experiments.

\noindent (3) The annihilation cross section of $e^+e^-$ through
$J/\Psi$ is about two order of magnitude larger than that without going
through $J/\Psi$ and the
branching ratios for our interested channels
from $J/\Psi$ decay are quite large\cite{PDG}, cf. Table~\ref{table:1}.

\vspace{-0.3cm}
\begin{table}[htb]
\caption{ Branching ratios (BR$\times 10^3$) for some interested channels}
\label{table:1}
\renewcommand{\arraystretch}{1.2} 
\begin{center}
\begin{tabular}{cccc}
\hline
$p\bar n\pi^-$  & $p\bar p\pi^0$  &  $p\bar p\pi^+\pi^-$ &
$p\bar p\eta$ \\\hline
$2.0\pm 0.1$ & $1.1\pm 0.1$ & $6.0\pm 0.5$ & $2.1\pm 0.2$ \\\hline
 $p\bar p\eta'$ & $p\bar p\omega$ &
$\Lambda\bar\Sigma^-\pi^+$ & $pK^-\bar\Lambda$\\\hline
$0.9\pm 0.4$ & $1.3\pm 0.3$ & $1.1\pm 0.1$ & $0.9\pm 0.2$ \\
\hline
\end{tabular}\\
\end{center}
\end{table}

\noindent With present available 7.8 million $J/\Psi$ events at BES-I and
forthcoming 50 millions more at BES-II in near future,
we expect to have enough statistics to get the best determination of
properties for $N^*\to\pi N$
and $\pi\pi N$ from $p\bar n\pi^-$, $p\bar p\pi^0$ and  $p\bar
p\pi^+\pi^-$ channels;
We can also search for the ``missing" $N^*$ states and study known $N^*$
states decaying into $\eta N$, $\eta' N$, $\omega N$ and $K\Lambda$ etc.

\noindent (4) On theoretical side, the process $J/\Psi\to\bar pN^*$ or
$p\bar N^*$ provides a
new way to probe the internal quark-gluon structure of the $N^*$
resonances\cite{Zou1}.

\vspace{-1.6cm}
\begin{figure}[htb]
\epsfig{figure=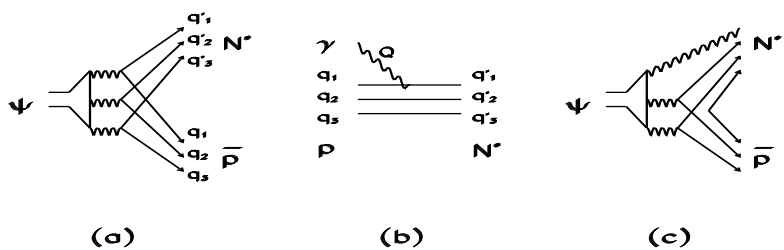,height=2.5in,width=5.4in}
\vspace{-1.6cm}\caption{Diagrams for $N^*$ production from $J/\Psi$
decays for 3-quark (qqq) $N^*$ (a) and hybrid (qqqg) $N^*$ (c), and
from $\gamma N$ reaction (b).}
\label{figure:2}
\end{figure}

In the simple three-quark picture of baryons, the process can be described
by Fig.~\ref{figure:2}(a). In this picture, three quark-antiquark pairs
are created independently via a symmetric three-gluon intermediate state;
the quarks and antiquarks have momenta of very similar magnitude.
This is quite different from the $\gamma p\to N^*$ process, cf.
Fig.~\ref{figure:2}(b),
where the photon couples to only one quark and unsymmetric configuration
of quarks is favored with $q'_1=q_1+Q$.   Therefore the processes
$J/\Psi\to\bar pN^*$ and
$\gamma p\to N^*$ should probe different aspects of the quark
distributions inside baryons. This may help us to distinguish various
quark models\cite{Isgur} 

For a hybrid $N^*$, since the $J/\Psi$ decay is a glue-rich process,
it can be produced via diagram Fig.~\ref{figure:2}(c) and is
expected to have larger production rate than a pure three-quark
$N^*$ resonance\cite{Page}.

Considering these advantages, a $N^*$ program at BEPC has been
proposed\cite{Zou2} and started\cite{Lihb}.

\section{Status of $N^*$ data at BES}

Based on 7.8 million $J/\Psi$ events collected at BEPC before 1996,
the events for $J/\Psi\to\bar pp\pi^0$ and $\bar pp\eta$ have been
selected and reconstructed\cite{Lihb}.  

\begin{figure}[htbp] 
\begin{minipage}[t]{55mm}
\centerline{\epsfig{file=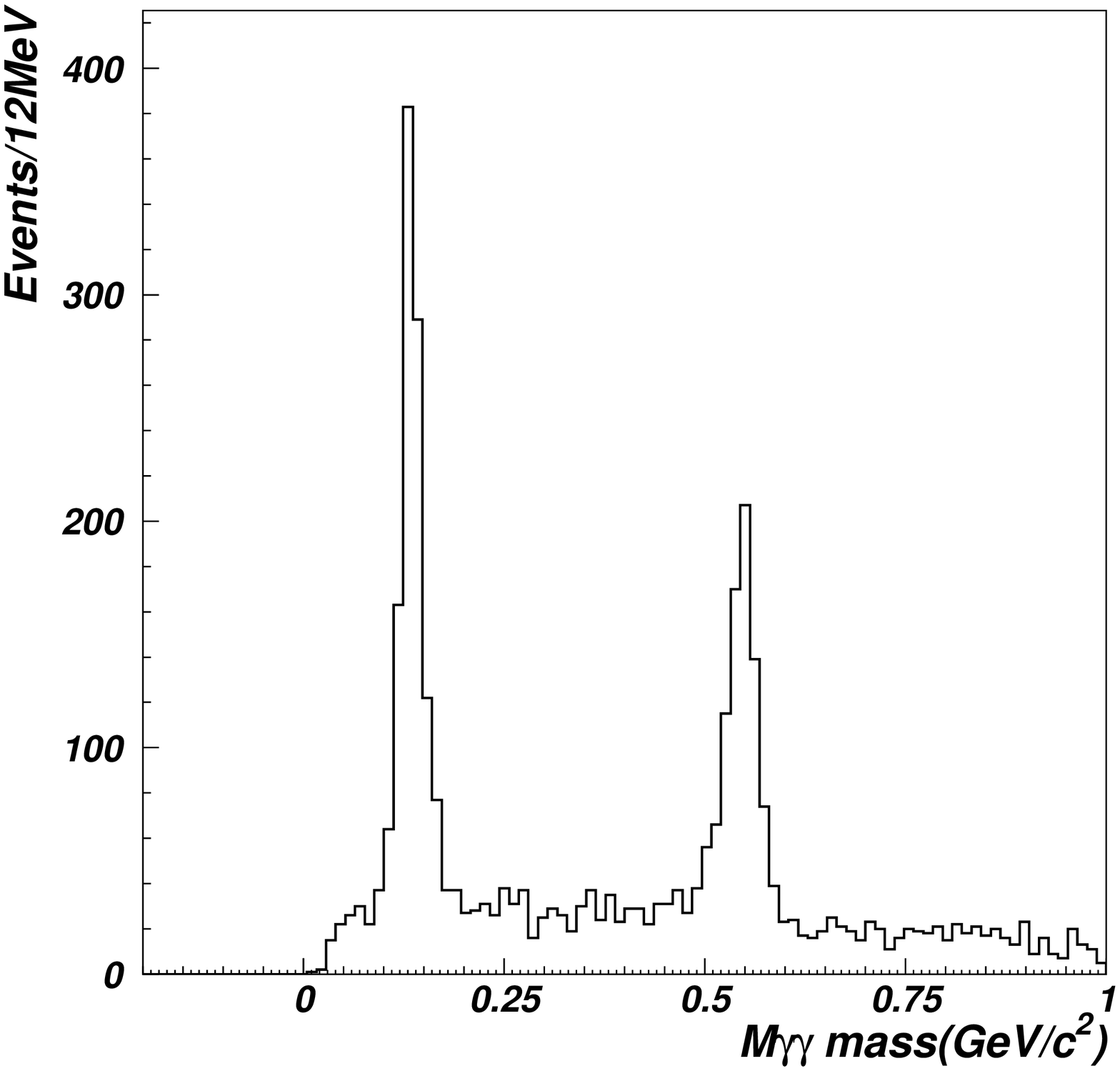,height=2.5in,width=2.5in}}
\caption[]{$\gamma \gamma$ invariant mass spectrum after 4C fit
for
$J/\Psi\to\bar pp\gamma\gamma$}
\label{figure:3}
\end{minipage}
\hspace{\fill}
\begin{minipage}[t]{55mm}
\centerline{\epsfig{file=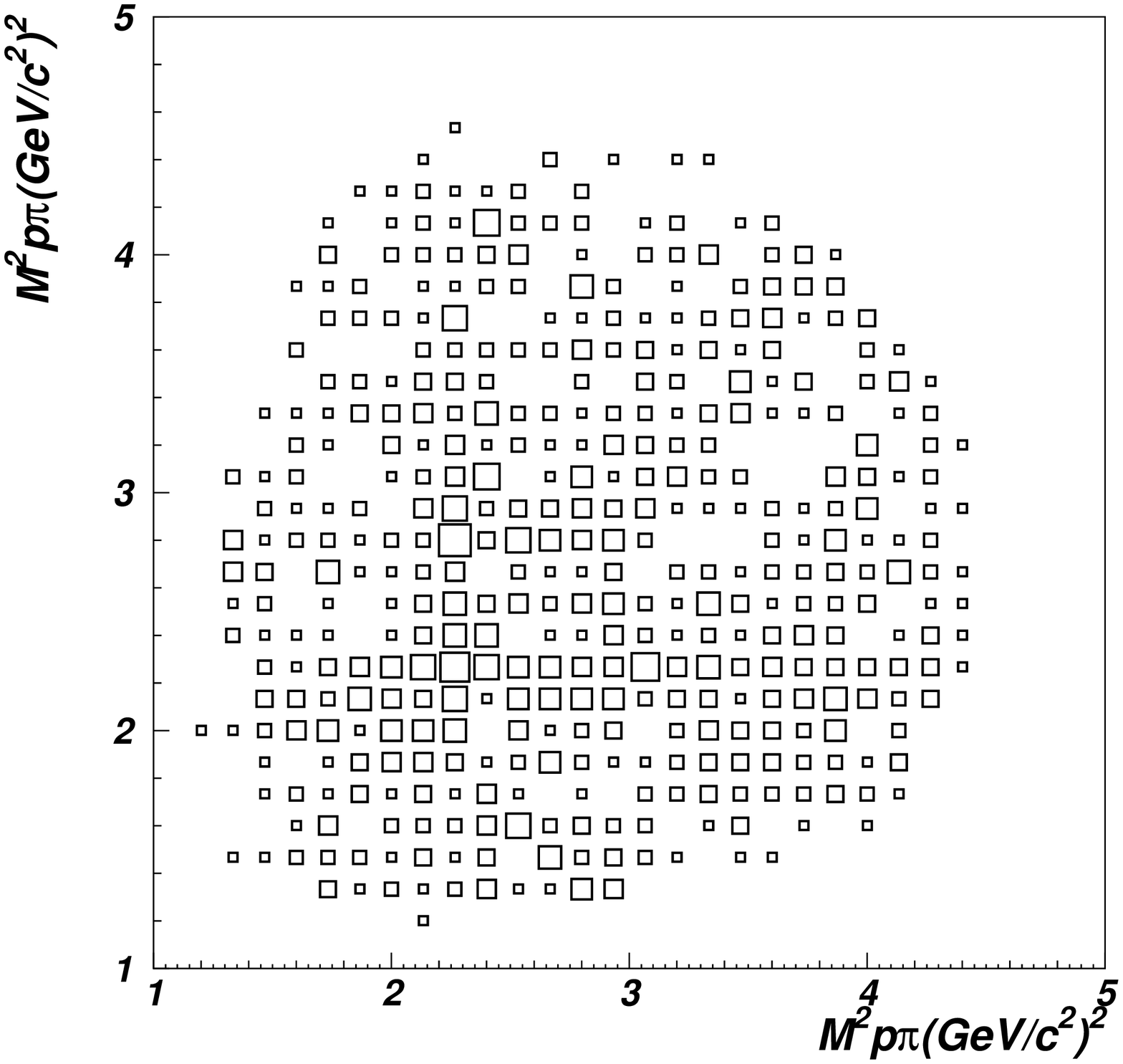,height=2.5in,width=2.5in}}
\caption{Dalitz plot for $J/\Psi\to\bar pp\pi^0$.}
\label{figure:4}
\end{minipage}
\end{figure}

The $\pi^0$ and $\eta$ are detected in their $\gamma\gamma$ decay mode.
For selected $J/\Psi\to\bar pp\gamma\gamma$ events, the invariant mass
spectrum of the $2\gamma$ is shown in Fig.\ref{figure:3}.
The $\pi^0$ and $\eta$ signals are clearly there.

\begin{figure}[htbp] 
\begin{minipage}[t]{55mm}
\centerline{\epsfig{file=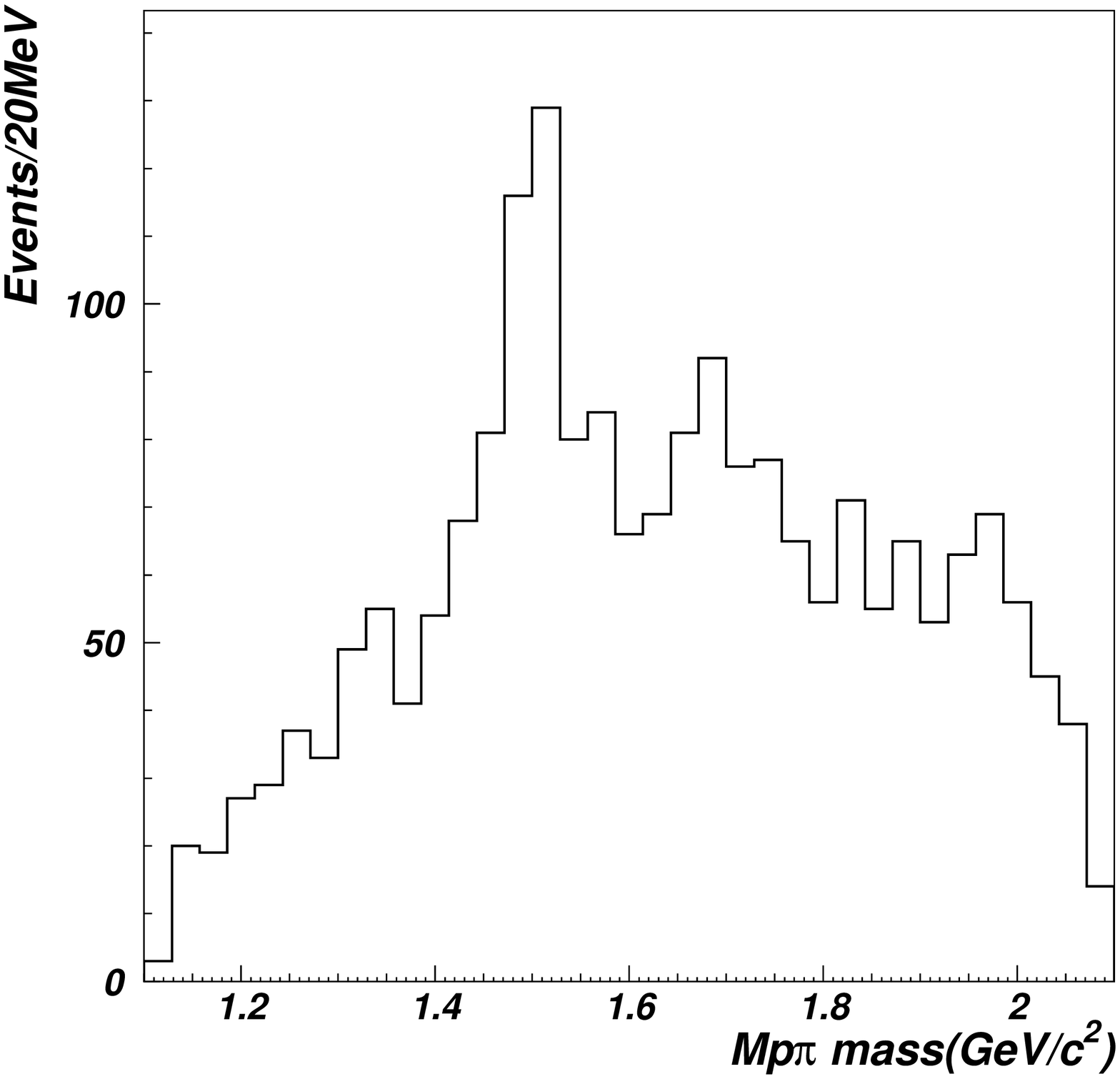,height=2.5in,width=2.5in}}
\caption[]{$p\pi^0$ invariant mass spectrum for $J/\Psi\to\bar
pp\pi^0$.}
\label{figure:5}
\end{minipage}
\hspace{\fill}
\begin{minipage}[t]{55mm}
\centerline{\epsfig{file=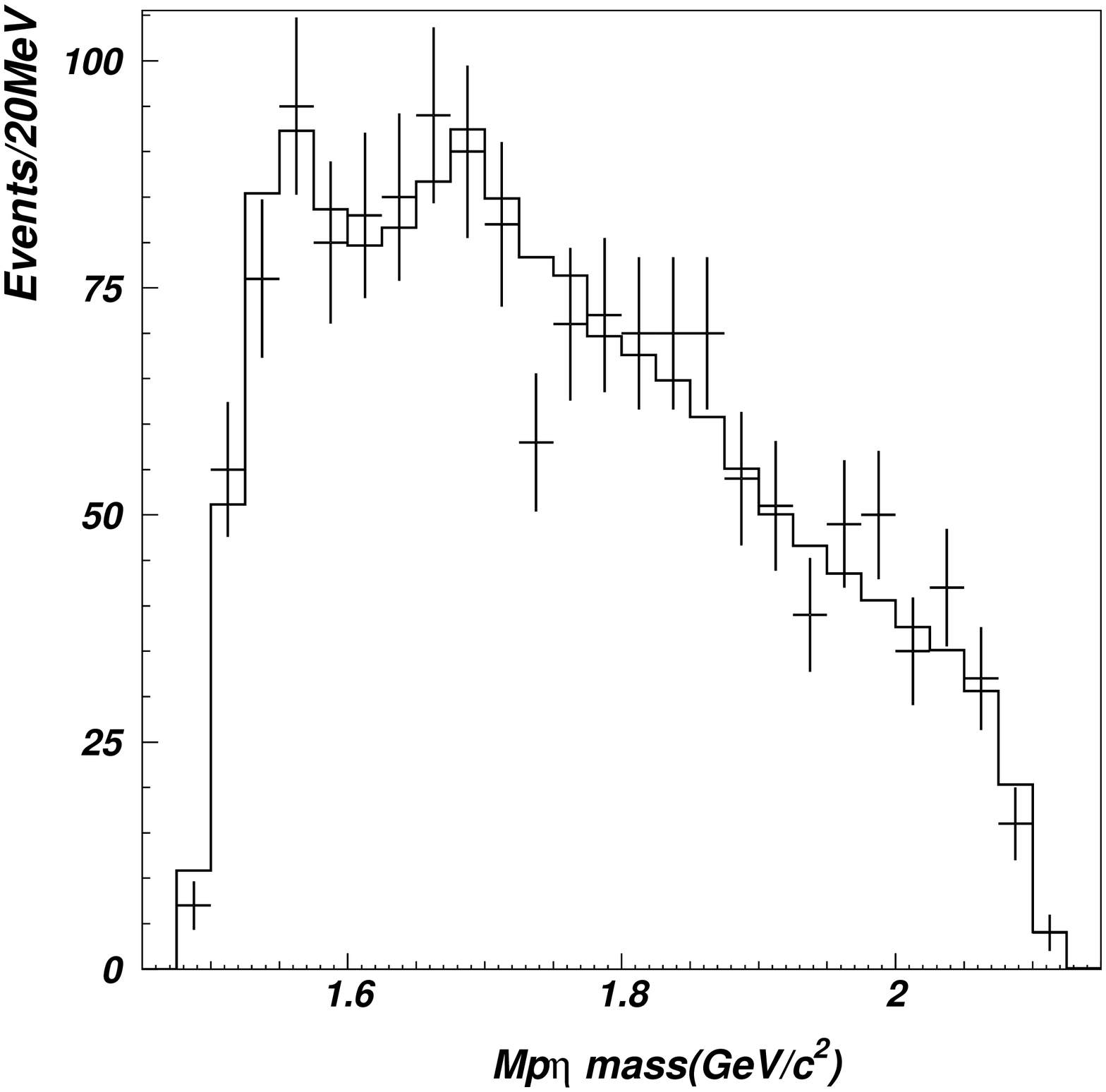,height=2.5in,width=2.5in}}
\caption{$p\eta$ invariant mass spectrum for $J/\Psi\to\bar pp\eta$,
Crosses are data and histogram the
fit}
\label{figure:6}
\end{minipage}
\end{figure}   

Fig.\ref{figure:4} and Fig.\ref{figure:5} show Dalitz plot and
$p\pi^0$ invariant mass spectrum for the $J/\Psi\to\bar pp\pi^0$.
There are clear peaks around 1480 and 1650 MeV of $p\pi^0$
invariant mass.

Fig.\ref{figure:6} shows $p\eta$ invariant mass spectrum for
the $J/\Psi\to\bar pp\eta$. There are clear enhancement around
the $p\eta$ threshold, peaks at 1540 and 1650 MeV.

We have also selected and reconstructed events for $J/\Psi\to\bar pn\pi^+$
and $p\bar n\pi^-$ channels. There are similar $p\pi$ structures as 
in $J/\Psi\to\bar pp\pi^0$ process.
The data processing for other channels, such as $\bar p\Lambda K$,
$\bar\Lambda pK$, $\bar pp\pi^+\pi^-$ and $\bar pp\omega$, are in
progress.

\section{Partial wave analyses of $J/\Psi\to\bar pp\eta$}

Because the $J/\Psi\to\bar pp\eta$ has the simplest possible resonance
contributions, a partial wave analysis is performed for this channel
first. We are mainly interested in the structures at 1540 and 1650 MeV of
the $p\eta$ invariant mass. 
Only $J^P={1\over 2}^\pm$ $N^*$ resonances are included in the
analysis, since according to the information\cite{PDG,Svarc} from $\pi
N\to\eta N$ and $\gamma N\to\eta N$ experiments resonances with higher
spins have much smaller couplings to $p\eta$ in our interested mass range.

We use the effective Lagrangian approach\cite{Nimai,Olsson} for the
partial wave analysis. The relavent spin-1/2 interaction Lagrangians are
\bea
{\cal L}_{\eta PR} &=& -ig_{\eta PR}\bar P\Gamma R\eta + H.c. , \\
{\cal L}^{(1)}_{\Psi PR} &=& \frac{ig_{T_R}}{M_R+M_P}\bar R\Gamma_{\mu\nu}
q^\nu P\Psi^\mu +H.c. , \\
{\cal L}^{(2)}_{\Psi PR} &=& -g_{V_R}\bar R\Gamma_\mu P\Psi^\mu +H.c. 
\eea
where $R$ is the generic notation for the resonance with mass $M_R$, $P$
for proton with mass $M_P$ and $\Psi$ for $J/\Psi$ with four-momentum $q$.
The operator structures for the $\Gamma$, $\Gamma_\mu$ and
$\Gamma_{\mu\nu}$ are
\be
\Gamma=1, \quad \Gamma_\mu=\gamma_\mu, \quad
\Gamma_{\mu\nu}=\gamma_5\sigma_{\mu\nu},
\label{eq:4}
\ee
\be
\Gamma=\gamma_5, \quad \Gamma_\mu=\gamma_5\gamma_\mu, \quad
\Gamma_{\mu\nu}=\sigma_{\mu\nu},
\label{eq:5}
\ee
where (\ref{eq:4}) and (\ref{eq:5}) correspond to nucleon resonances of
odd and even parities, respectively. The relative magnitudes and phases
of the amplitudes are determined by a maximum likelihood fit to the data.
A fit with three $N^*$ resonances is shown in Figs.6-8 for the
$p\eta$, $\bar pp$ invariant mass spectra and the angular distribution
of the proton relative to the beam direction, respectively.

\begin{figure}[htbp] 
\begin{minipage}[t]{55mm}
\centerline{\epsfig{file=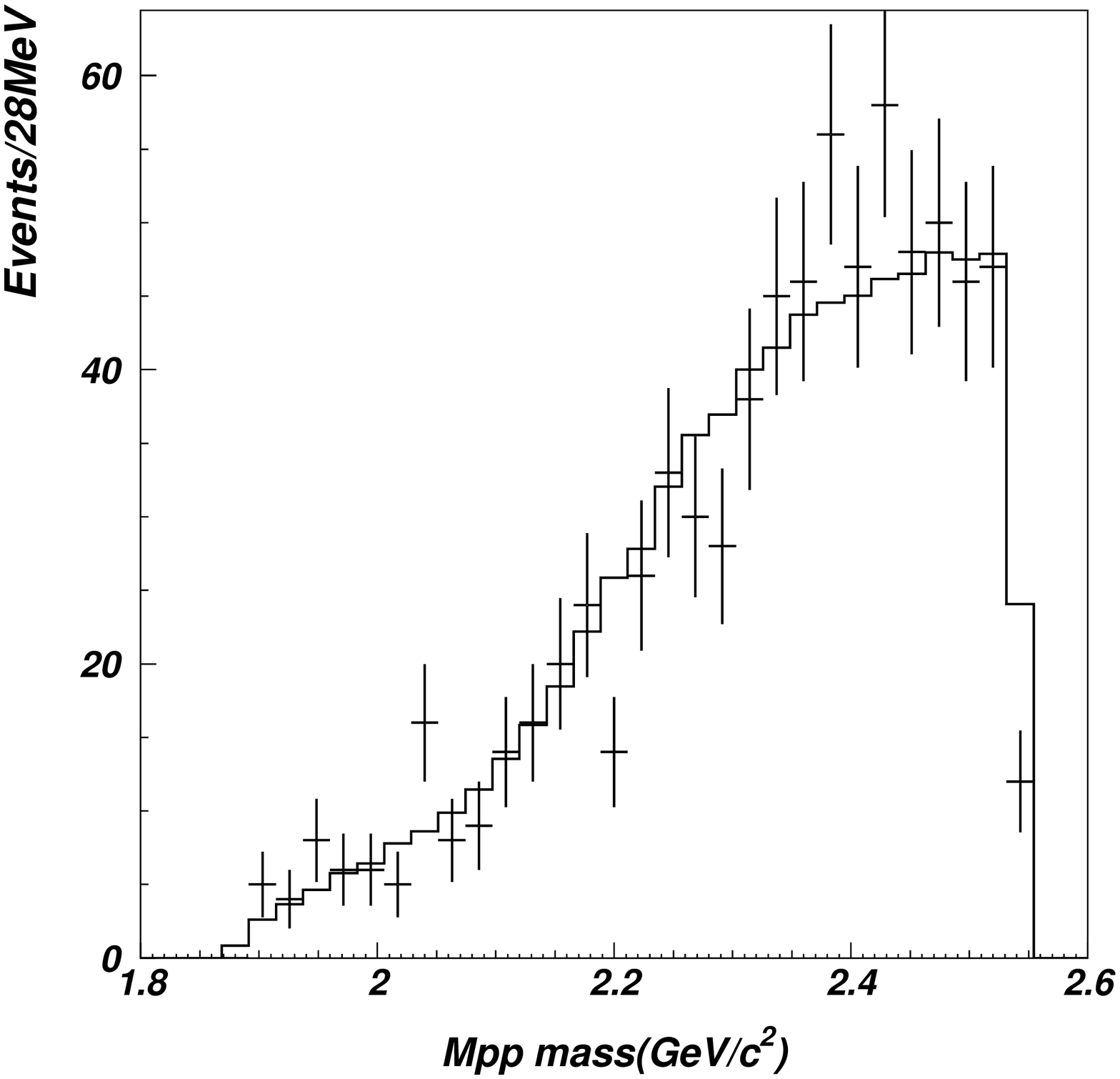,height=2.5in,width=2.5in}}
\caption[]{$p\bar p$ invariant mass spectrum for $J/\Psi\to\bar
pp\eta$. Crosses are data and histogram the fit.}
\label{figure:7}
\end{minipage}
\hspace{\fill}
\begin{minipage}[t]{55mm}
\centerline{\epsfig{file=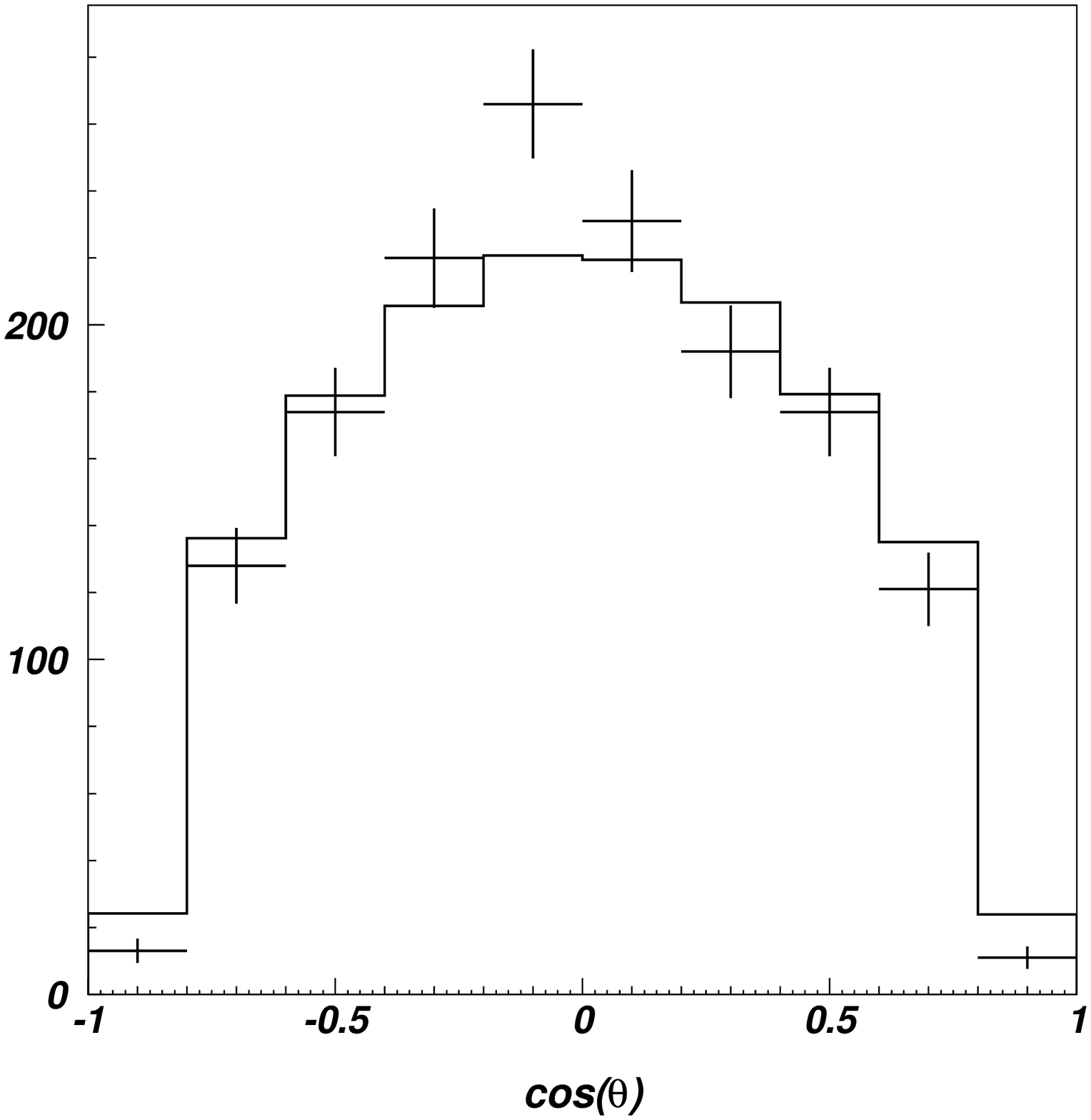,height=2.5in,width=2.5in}}
\caption{Angular distribution of proton relative to the beam direction.
Crosses are data and histogram the fit.}
\label{figure:8}
\end{minipage}
\end{figure}   

The peak near the $p\eta$ threshold is fitted with a $N^*$ resonance
of mass and width optimised at $M = 1540^{+15}_{-17}$ and $\Gamma =
178^{+20}_{-22}$ MeV, respectively. The data favor  $J^{P} =
\frac{1}{2}^{-}$ over $ \frac{1}{2}^{+}$. A fit with $J^{P} =
\frac{1}{2}^{+}$ instead gives likelihood value $lnL$ worse by 16.0
than for $\frac{1}{2}^{-}$ assignment (Our definition of $lnL$ is such
that it increases by 0.5 for a one standard deviation change in one
parameter). The statistical significance of the peak is above 6.0$\sigma$.
It is obviously the $S11$ $N^*(1535)$ resonance. It makes the largest
contribution $(84\pm 5)\%$ to the $p\bar p\eta$ final states. The errors
here and later include both statistics and systematic errors from the fit.

The second peak around 1650 MeV is also fitted with a $J^{P}=\frac{1}{2}^-$ 
resonance $N^*(1650)$. Its mass optimises at $M=1648^{+18}_{-16}$ MeV with
$\Gamma = 150$ MeV fixed to PDG value. Its width cannot be well determined
by our data due to a correlation with the parameters used for the third
resonance. It contributes $(11\pm 3)\%$ to the $p\bar p\eta$ final
states.

A small improvement to the fit is given by including an addtional
$J^{P}=\frac{1}{2}^+$ resonance, which optimises at $M=1834^{+46}_{-55}$
MeV with $\Gamma = 200$ MeV fixed. The statistical significance of the
peak is only $2.0 \sigma$. We have tried $J^{P}=\frac{1}{2}^{-}$ instead
$\frac{1}{2}^+$, the fit is worse.

An interesting result is that the ${\cal L}^{(2)}_{\Psi PR}$ term given
by Eq.(3) makes insignificant contribution for both $N^*(1535)$
and $N^*(1650)$. If we drop this kind of couplings for both resonances,
the likelihood value for the fit is only worse by 0.8 for 4 less
free parameters. This kind of couplings should vanish for the real
photon coupling to $NN^*$. Why it also vanishes for the $\Psi NN^*$
coupling needs to be understood. A theoretical calculation\cite{Okubo}
assuming pure ${\cal L}^{(2)}_{\Psi PR}$ coupling without ${\cal
L}^{(1)}_{\Psi PR}$ coupling failed to reproduce the basic feature
of the $J/\Psi\to\bar pp\eta$ data. This is consistent with our
observation that the ${\cal L}^{(1)}_{\Psi PR}$ coupling dominates
for both $N^*(1535)$ and $N^*(1650)$. 

In the Vector Meson Dominance (VMD) picture, the virtual photon couples
to the $NN^*$ through vetor mesons, and the electro-magnetic $NN^*$ 
transition form factors $g_{\gamma^*NN^*}$ can be expressed in terms of
photon-meson coupling strengths $C_{\gamma V}$ and meson-$NN^*$ vertex
form factors $g_{_{VNN^*}}$:
\be
g_{\gamma^*NN^*}(q^2)=\sum_j\frac{m_j^2C_{\gamma
V_j}}{m^2_j-q^2-im_j\Gamma_j}g_{_{V_jNN^*}}(q^2)
\ee
with 
\be
C_{\gamma V}=\sqrt{\frac{3\Gamma_{V\to e^+e^-}}{\alpha m_V}}.
\ee
At $q^2=M^2_\Psi$, the $J/\Psi$ meson dominates. The terms from other
vector mesons are negligible.
From our PWA results here and other relavent information from
PDG\cite{PDG}, we can deduce the transition form factor for the time-like
virtual photon to $PN^*(1535)$ as 
\be
|g_{\gamma^*pN^*}(q^2=M_\Psi^2)|= 2.8\pm 0.5 ,
\ee 
which is related to the more familiar helicity amplitude $A^P_{1/2}$ for
$N^*\to\gamma P$ by
\be
|A^P_{1/2}|^2=\left(\frac{g_{\gamma pN^*}(q^2=0)}{M_{N^*}+M_P}\right)^2
\frac{(M^2_{N^*}-M^2_P)}{2M_P} .
\ee

\section{Summary and outlook}

In summary, the $J/\Psi$ decay at BEPC provides a new excellent
laboratory for studying the $N^*$ resonances. On experimental side, 
it provides a natural isospin 1/2 filter for $\pi N$ and $\pi\pi N$
systems and many interesting channels for studying $N^*$ and hyperon
resonances;
on theoretical side, it provides a new way to explore the
internal structure of baryons and may help us to pin down hybrid
baryon(s).
Almost all subjects on the $N^*$ resonances at the CEBAF\cite{Burkert} and
other $\gamma p(ep)$ facilities can be studied here complementally with
the virtual time-like photon.

Based on 7.8 million $J/\psi$ events collected at BEPC, a partial wave
analysis of $J/\psi \to p\bar{p}\eta$ data has been performed. Two 
$J^P=\frac{1}{2}^-$ resonances, $N^*(1535,S_{11})$ and $N^*(1650,S_{11})$,
are observed. Now we are collecting more $J/\Psi$ events with the
improved BES detector. With the
forthcoming 50 millions more $J/\Psi$ events in near future, more
precise partial wave analyses can be carried out on many channels
involving $N^*$ resonances and should offer some best determinations of
$N^*$ properties. A systematic theoretical and experimental  
study of the $N^*$ and hyperon production from the $J/\psi$ decays
is underway.

There is also a plan to upgrade the BEPC to BEPC2 which will increase the
luminosity by an order of magnitude. This will provide us more precise
data for the study of the $N^*$ resonances from $J/\Psi$ decays and also
enable us to  extend the $N^*$ program to a higher energy at the $\Psi'$
resonance which now suffers low statistics for the $N^*$ study.

With the new generation of $\gamma p(ep)$, $J/\psi$ and $\Psi'$
experiments, a new era for the baryon spectroscopy is coming.

\eject


\vspace*{-9pt}

\section*{References}
\begin{thebibliography}{99}
\bibitem{PDG} Particle Data Group, \Journal{\EPJC}{3}{1}{1998}.
\bibitem{Workman} R.Workman, {\em Few Body Syst. Suppl.} {\bf 11}, 94 (1999).
\bibitem{Zou1} B.S.Zou, G.X.Peng, H.C.Chiang and P.N.Shen, hep-ph/9909204.
\bibitem{Isgur} N.Isgur and G.Karl, \Journal{\PRD}{18}{4187}{1978};
S.Capstick and N.Isgur, \Journal{\PRD}{34}{2809}{1986};
R.Bijker et al., \Journal{\em Ann. Phys.}{236}{69}{1994};
K.F.Liu and C.W.Wong, \Journal{\PRD}{28}{170}{1983};
L.Ya.Glozman and D.O.Riska, \Journal{\em Phys. Rep.}{268}{1}{1996};
P.N.Shen et al., \Journal{\PRC}{53}{2024}{1997};
N.Kaiser, T.Waas and W.Weise, \Journal{\NPA}{612}{297}{1997};
Y.B.Dong et al., \Journal{\PRC}{60}{035203}{1999};
T.Barnes and F.E.Close, \Journal{\PLB}{123}{89}{1983}.
\bibitem{Page} S.Capstick and P.Page, \Journal{\PRD}{60}{111501}{1999}.
\bibitem{Zou2} B.S.Zou, Talks at CCAST workshop on BES Future Physics,
Aug. 1998,  and SLAC Workshop on $\tau-C$ Physics, March 1999. 
\bibitem{Lihb} BES Collaboration (H.B.Li et al.), Preprint hep-ex/9910032,
to be published in Nucl. Phys. A.
\bibitem{Svarc} M.Batinic, I.Slaus and A.Svarc,
\Journal{\PRC}{51}{2310}{1995}; 
A.M.Green and S.Wycech, \Journal{\PRC}{60}{035208}{1999}.  
\bibitem{Nimai} M.Benmerrouche, N.C.Mukhopadhyay and J.F.Zhang, 
\Journal{\PRL}{77}{4716}{1996}; \Journal{\PRD}{51}{3237}{1995}.
\bibitem{Olsson} M.G.Olsson and E.T.Osypowski,
\Journal{\NPB}{87}{399}{1975}; \Journal{\PRD}{17}{174}{1978};
M.G.Olsson et al., \Journal{\em ibid.}{17}{2938}{1978}.
\bibitem{Okubo} R. Sinha and S. Okubo, \Journal{\PRD}{30}{2333}{1984}. 
\bibitem{Burkert} V.D.Burkert, \Journal{\NPA}{623}{59c}{1997}.
\end{thebibliography}
\end{document}